\documentclass[a4paper,11pt]{article}
\usepackage{pos}

\title{Collective neutrino oscillations accounting for neutrino quantum decoherence}

\author*[a]{Konstantin Stankevich}
\author[a,b]{Alexander Studenikin}

\affiliation[a]{ Faculty of Physics, Lomonosov Moscow State University,\\ Moscow 119991, Russia}

\affiliation[b] {Joint Institute for Nuclear Research,\\ Dubna 141980, Moscow Region, Russia}

\emailAdd{kl.stankevich@physics.msu.ru}
\emailAdd{studenik@srd.sinp.msu.ru}

\abstract{
	
In our previous studies (see \cite{Stankevich:2020icp} and references therein)  we developed a new theoretical framework that enabled one to consider a new mechanism of neutrino quantum decoherence engendered by the neutrino radiative decay. In parallel, another framework was developed (see \cite{Nieves:2020jjg} and references therein) for the description of the neutrino quantum decoherence due to the non-forward neutrino scattering processes. Both mechanisms are described by the master equations in the Lindblad form.

In the present studies we are are not interested in a specific mechanism of neutrino quantum decoherence. Therefore, we use the general Lindblad master equation for the description of the neutrino quantum decoherence and do not fix an analytical expressions for the decoherence and relaxation parameters. 
	
We study the influence of the neutrino quantum decoherence on collective neutrino oscillations. Collective neutrino oscillations is a phenomenon engendered by neutrino-neutrino interaction. It is significant in different astrophysical environments where the neutrino density is extremely high. Examples of such environments are the early universe, supernovae explosions, binary neutron stars, accretion discs of black holes.  The effect of collective neutrino oscillations attracts the growing interest in sight of appearance of multi-messenger astronomy and constructing of new large-volume neutrino detectors that will be highly efficient for observing neutrino fluxes from supernovae explosions. Previously, it was shown that neutrino quantum decoherence can significantly modify the neutrino fluxes from reactors and the sun. Here below we study the influence of the neutrino quantum decoherence on supernovae neutrino fluxes. The peculiarity of the supernovae fluxes is that one of the main modes of neutrino oscillations in supernovae is engendered by the collective effects. Note, in the previous studies dedicated to the collective neutrino oscillations only the kinematical decoherence (see \cite{Capozzi:2019lso} and references therein) was accounted for.
}

\FullConference{%
  40th International Conference on High Energy physics - ICHEP2020\\
  July 28 - August 6, 2020\\
  Prague, Czech Republic (virtual meeting)
}

\begin{document}
\maketitle

Consider the two-flavor neutrino mixing scenarios, i.e. the mixing between $\nu_e$ and $\nu_x$ states where $\nu_x$ stands for $\nu_\mu$ or $\nu_\tau$.
Here below we focus on the derivation of the neutrino oscillation probability and highlight the interplay between collective neutrino oscillations and neutrino quantum decoherence. We use the simplified model of supernova neutrinos that was considered in \cite{Vaananen:2015hfa, Stapleford:2016jgz}. In such a model neutrinos are produced and emitted with a single energy and a single emission angle.

The neutrino evolution in supernovae environment that accounts for neutrino quantum decoherence is determined by the following equations

\begin{equation}
i \dfrac{d \rho_f}{d t} = [H, \rho_f] + D[\rho_f] \ , \ \ \ \ \ \ \ \ i \dfrac{d \bar{\rho}_f}{d t} = [\bar{H}, \bar{\rho}_f] + D[\bar{\rho}_f]
\label{Lindblad}
,
\end{equation}
where $\rho_f$ ($\bar{\rho}_f$) is the density matrix for neutrino (antineutrino) in the flavour basis and $H$ ($\bar{H}$) are total neutrino (antineutrino) Hamiltonian. Neutrino quantum decoherence is described by the dissipation term $D[\rho]$ that we define in the next section.

Hamiltonian $H$ contains the three terms $H = H_{vac} + H_{M} + H_{\nu \nu}$, where $H_{vac}$ is the vacuum Hamiltonian, $H_{M}$ and $H_{\nu \nu}$ are Hamiltonians that describe matter potential and neutrino-neutrino interaction correspondingly. The exact expressions one can find in \cite{Vaananen:2015hfa, Stapleford:2016jgz}. The dissipation term $D[\rho]$ we write in the Lindblad form 

\begin{equation}
D \left[ \rho_{\tilde{m}}(t) \right] = \dfrac 1 2 \sum^{3}_{k=1} \left[ V_k, \rho_{\tilde{m}} V^\dag_k \right] + \left[ V_k \rho_{\tilde{m}}, V_k^\dag \right]
,
\end{equation}
where $V_k$ are the dissipative operators that arise from interaction between the neutrino system and the external environment, $\rho_{\tilde{m}}$ is the neutrino density matrix in the effective mass basis. Here below, we  omit index ``$\tilde{m}$'' in order not overload formulas. The operators $V_k$, $\rho_f$ and $H$ can be expanded over the Pauli matrices $O = a_\mu \sigma_\mu$, where $\sigma_\mu$ are composed by an identity matrix and the Pauli matrices. In this case eq. (\ref{Lindblad}) can be written in the following form

\begin{equation}
\dfrac{\partial P_k (t)}{\partial t} \sigma_k = 2 \epsilon_{ijk} H_i P_j (t) \sigma_k + D_{kl} P_l(t) \sigma_k
\label{LindbladEq}
,
\end{equation}
where the matrix $D_{ll} = - diag \{ \Gamma_1, \Gamma_1, \Gamma_2 \}$ and $\Gamma_1$, $\Gamma_2$ are the parameters that describe two dissipative effects: 1) the decoherence effect and 2) the relaxation effect, correspondingly. In the case of the energy conservation in the neutrino system  there is an additional requirement on a dissipative operators \cite{Oliveira:2016asf} $[H_S, V_k]=0$. In this case the relaxation parameter is equal to zero $\Gamma_2=0$. Here below, we consider only the case of the energy conservation, i.e. $\Gamma_2=0$.

Here below, we consider analytical conditions for the occurrence of the neutrino collective effects. The onset of these collective effects has been related to the presence of an instability (see \cite{Vaananen:2015hfa} and references therein). In order to study this instabilities we will apply to eq. (\ref{Lindblad}) the linearization procedure described in \cite{Vaananen:2015hfa, Vaananen:2013qja}. Consider a time dependent small amplitude variation $\delta P_k$ around the initial configuration $P^0_k$ and a corresponding variation of the density dependent Hamiltonian $\delta H_k$ around the initial Hamiltonian $H^0_k$: $P_k = P_k^0 +\delta P_k$ where $\delta P_k = P_k' e ^{-i \omega t} + H.c.$ and $H_k = H_k^0 +\delta H_k$ where $\delta H_k = H_k' e ^{-i \omega t} + H.c.$. Where one can write $H_k' = \dfrac{\partial H_k}{\partial P_k} P_k' + \dfrac{\partial H_k}{\partial \bar P_k} \bar P_k'$. In the case of high electron density the in-medium eigenstates initially coincide with the flavor states. Therefore, the initial conditions are given by $H_k^0 = \left( 0, 0, H^0 \right)^T$ and $P_k^0 = \left( 0, 0, P^0 \right)^T$. Putting everything together and considering only the non-diagonal elements ($\rho_{12} = P_x + i P_y $) one obtains the following equation for eigenvalues  (we neglect the higher-order corrections)

\begin{equation}
(\omega - i \Gamma_1)
\left(
\begin{matrix}
\rho_{12}' \\
\bar \rho_{21}'
\end{matrix}
\right)
=
\left(
\begin{matrix}
A_{12} & B_{12} \\
\bar A_{21} & \bar B_{21}
\end{matrix}
\right)
\left(
\begin{matrix}
\rho_{12}' \\
\bar \rho_{21}'
\end{matrix}
\right)
,
\end{equation}
where on the right-hand side of equation is the stability matrix that coincides with one from \cite{Vaananen:2015hfa, Vaananen:2013qja}. In case of a single energy and single emission angle it is expressed as

\begin{equation}
\begin{matrix}
A_{12} = (H^0_{11} - H^0_{22}) - \dfrac{\partial H_{12}}{\partial \rho_{12}} (\rho_{11}^0-\rho_{22}^0), \ \ \ \ \ \ \ \ \
B_{12} = \dfrac{\partial H_{12}}{\partial \bar \rho_{21}^0} (\rho_{22}^0 - \rho_{11}^0), \\
\bar A_{21} =  (\bar H^0_{22} - \bar H^0_{11}) - \dfrac{\partial \bar H_{21}}{\partial \bar \rho_{21}} (\bar \rho_{22}^0-\bar \rho_{11}^0), \ \ \ \ \ \ \ \ \
\bar B_{21} = \dfrac{\partial \bar H_{21}}{\partial \rho_{12}^0} (\bar \rho_{11}^0 - \bar \rho_{22}^0).
\end{matrix}
\end{equation}

The eigenvalues are given by $\omega = i\Gamma_1 + \dfrac 1 2 \left( A_{12} + \bar A_{21} \pm \sqrt{(A_{12}-\bar A_{21})^2 +4 B_{12} \bar B_{21} }   \right)$. One can see that if the eigenvalues have an imaginary part, the non-diagonal elements of the neutrino density matrix can grow exponentially and thus the system become unstable, that is, if

\begin{equation}\label{conditions}
\begin{cases}
(A_{12}-\bar A_{21})^2 +4 B_{12} \bar B_{21}  < 0, \\
\mathcal{I}m \left( (A_{12}-\bar A_{21})^2 +4 B_{12} \bar B_{21} \right)  > \Gamma_1
.
\end{cases}
\end{equation}

The first condition is the same as was derived in \cite{Vaananen:2015hfa, Vaananen:2013qja}. The second term is a new one that was not considered before. From eq. (\ref{conditions}) one can see, that neutrino quantum decoherence prevents a system from an exponential growth of non-diagonal elements, i.e. neutrino quantum decoherence leads to the damping of the neutrino collective oscillations.

In this paper we considered for the first time the effect of the neutrino quantum decoherence in supernovae fluxes. We derive  new conditions of the collective neutrino oscillations accounting for neutrino quantum decoherence which can appear as a result of the physics beyond the Standard Model. This work was supported by the Russian Foundation for Basic Research under Grant No. 20-52-53022-GFEN-a. The work of KS was also supported by the Russian Foundation for Basic Research under Grant No. 20-32-90107.

\end{document}